\documentclass{kapedbk} 
\usepackage{epsfig}
\let\footnote\savefootnote
\let\footnotetext\savefootnotetext 
\let\footnoterule\savefootnoterule 
\kluwerbib
\normallatexbib
\begin{document}
\articletitle[]{Transport in one dimensional\\ quantum systems}
\author{X. Zotos}
\affil{University of Fribourg\\
and IRRMA (EPFL - PPH), 1015 Ecublens, Switzerland}
\email{Xenophon.Zotos@epfl.ch}
\author{P. Prelov\v sek}
\affil{Faculty of Mathematics and Physics, University of Ljubljana,\\
and J. Stefan Institute, 1000 Ljubljana, Slovenia}
\email{peter.prelovsek@ijs.si}
\begin{abstract}
In this Chapter, we present recent theoretical developments 
on the finite temperature transport of one dimensional electronic 
and magnetic quantum systems as described by a variety of prototype models. 
In particular, we discuss the unconventional transport and dynamic - spin, 
electrical, thermal - properties implied by the integrability of models as 
the spin-1/2 Heisenberg chain or Hubbard. 
Furthermore, we address the implication of these 
developments to experimental studies and theoretical descriptions 
by low energy effective theories.

\end{abstract}
\begin{keywords}
one dimensional quantum many body systems,\\transport, integrability
\end{keywords}
\section{Introduction}
The electronic and magnetic properties of reduced dimensionality materials
are significantly modified by strong correlation effects.
In particular, over the last few years, the physics of  quasi-one 
dimensional electronic systems, has been the focus of an ever increasing 
number of theoretical and experimental studies. They are realized as 
three dimensional (3D) compounds composed of weakly
interacting chains or, in a latest and very promising development, 
as monoatomic width chains fabricated by self-assembly on surfaces. 

Experimentally, recent studies made possible by the synthesis of new
families of compounds characterized by very weak interchain coupling 
and low disorder, indicate unconventional transport and dynamic 
behavior; for example, unusually high thermal conductivity in 
quasi-one dimensional magnetic compounds \cite{ott,thcond1,thcond2}, 
ballistic spin transport in magnetic 
chains \cite{takigawa,thurber} or optical conductivity in quasi-1D 
organic conductors, showing a low frequency very narrow ``Drude peak" 
even at relatively high temperatures \cite{dg,degiorgi}.

Theoretically, it is well known that one dimensional (1D) systems of 
interacting electrons do not follow the phenomenological description of the
ordinary Landau Fermi liquids, but rather they are characterized by a
novel class of collective quantum states coined Luttinger 
liquids \cite{haldll}.
Furthermore, it has quite recently been realized that
several prototype models commonly used to describe 1D materials imply
ideal transport properties (dissipationless) even at high
temperatures. This phenomenon is the quantum analogue
of transport by nondecaying pulses (solitons) in 1D classical
nonlinear integrable systems \cite{osol}.

\bigskip
Of course, 1D electronic and magnetic systems have, since the sixties, 
been a favorite playground where theoretical ideas were confronted with 
experimental results on an ever improving quality and variety of quasi-1D
compounds. We are now in a position to claim reliable theoretical
analysis on the thermodynamics, quantum phase transitions and
spectral functions of prototype many body Hamiltonians used to
describe 1D materials. The tools at hand range from exact analytical
solutions (e.g. using the Bethe ansatz (BA) method) \cite{korepin}, 
the low energy Luttinger
liquid approach and powerful numerical simulation
techniques as the Quantum Monte Carlo (QMC) \cite{qmc} and Density
Matrix Renormalization Group (DMRG) \cite{dmrg} method.
In particular, the ground-state properties as well as the low-temperature 
behavior of the 
correlation functions and of most static quantities in the scaling
(universal) regime of Luttinger liquids have been extensively studied in 
recent decades and are well understood by 
now \cite{emery,solyom,voit,schoen}. 

On the other hand, although most experimentally relevant, 
less studied and understood are the transport and dynamic 
properties of 1D interacting electronic or magnetic systems. 
Paradoxically, while the
equilibrium properties of prototype integrable models as the spin-1/2
Heisenberg or Hubbard model have been extensively analyzed, their
finite temperature transport has attracted little attention; most
studies till a few years ago, have basically focused on the low energy
description in terms of the Luttinger model. A noticeable exception
has been the issue of diffusive versus ballistic behavior and thermal
conductivity of spin chains, a long standing and controversial issue.
The difficulties encountered with the transport quantities can be 
attributed to the
fact that the scattering and dissipation in clean 1D fermionic
systems are not dominated by low-energy processes and thus the transport
properties are not universal. 

\bigskip
Presently, transport properties are at the focus of intense theoretical
activity; in particular, prototype integrable models (as the Heisenberg,
t-J, Hubbard, nonlinear-$\sigma$) are studied by exact analytical
techniques (e.g. Bethe ansatz, form factor method) and numerical
simulations. However, the complexity of these methods often renders
the resulting behavior still controversial. Furthermore, the
transport of quasi-one dimensional systems is (re-) analyzed within
the effective Luttinger liquid theory or by semiclassical, Boltzmann type, 
approaches.

It is fair to say that the study of finite temperature/frequency 
conductivities in strongly correlated systems presents at the moment
fundamental conceptual as well as technical challenges.  
Development of new analytical and numerical simulation techniques
is required, as well as progress on the basic
understanding of scattering mechanisms and their effects.

\bigskip
In the following, we will mostly concentrate on the conductivity of
bulk, clean systems where the scattering mechanism is due to electronic or
magnetic interactions (Umklapp processes).  In particular, we will not
address issues on the transport of mesoscopic systems (e.g. nanowires,
nanotubes) or other dissipation mechanisms as coupling to
phonons or disorder.

In section 2 we start by presenting some elements of linear response 
theory (or Kubo formalism), the theoretical framework commonly used 
for describing transport properties. 
Then, in section 3 and 4 we continue with a presentation of 
the state of the transport properties of prototype systems, in particular 
the Heisenberg and Hubbard model. In section 5, we present a short overview of 
alternative approaches based on low energy effective field theories as 
the Luttinger liquid, sine-Gordon and quantum nonlinear-$\sigma$ models. 
Finally, in section 6, we close 
with a critical assessment of the present status and a discussion of open 
issues.

This presentation is definitely not an exhaustive account of 
theoretical studies on the transport properties of one dimensional 
quantum systems but it rather aims at presenting a coherent and 
self-contained view of some recent developments.

\section{Linear response theory}
 
In this section we introduce the basic definitions and concepts that will 
be used in the later development.
The framework of most transport studies is linear response theory
where the conductivities are given in terms of finite temperature $(T)$
dynamic correlations calculated at thermodynamic equilibrium \cite{millis}.
For instance, the real part of the electrical conductivity 
at frequency $\omega$ is given by the corresponding dynamic current 
correlation $\chi_{jj}(\omega)$, 

\begin{eqnarray}
\sigma'(\omega)&=&2\pi D \delta(\omega)+\sigma_{reg}(\omega)
\label{econd}\\
\sigma_{reg}(\omega)&=&\Re \frac{1}{i \omega}\chi_{jj}(\omega)\\
\chi_{jj}(\omega)&=&i\int_0^{+\infty}dt e^{izt} \langle [j(t),j]\rangle, 
~~~z=\omega+i\eta
\label{econd2}
\end{eqnarray}

\noindent
with $j$ the appropriate current operator.
In a spectral representation the conductivity is,
\begin{equation}
\sigma_{reg}(\omega)=\frac{1-e^{-\beta\omega}}{\omega}
\frac{\pi}{L}\sum_n p_n \sum_{m\neq n} 
\mid \langle n \mid j \mid m \rangle \mid ^2
\delta(\omega-(\epsilon_m - \epsilon_n)), 
\label{spcund}
\end{equation}
$\mid n\rangle, \epsilon_n$ denoting the eigenstates and
eigenvalues of the Hamiltonian,
$p_n$ the corresponding Boltzmann weights and $\beta$ the inverse temperature
(in the following we take $\hbar=\kappa_B=e=1$);
the $dc$ conductivity is given by the limit
$\sigma_{dc}=\sigma_{reg}(\omega\rightarrow 0)$.
We will mostly discuss one dimensional tight-binding models on $L$ sites 
where the current operator does not commute with the Hamiltonian.

To define the current operators we use the continuity equations  
of charge, magnetization or energy for the  
electrical, magnetic and thermal conductivity, respectively.
We will explicitly present them below in the discussion of the Heisenberg 
and Hubbard models.

\bigskip
A quantity that presently attracts particular attention is the prefactor $D$ 
of the $\delta-$function, named the Drude weight or charge stiffness.
This quantity was introduced by W. Kohn in 1964 as a criterion of (ideal) 
conducting or insulating behavior \cite{kohn} at $T=0$ 
in the context of the Mott-Hubbard transition. 
This meaning becomes clear by noting that $D$ 
is also the prefactor of the low frequency, imaginary   
(reactive - nondissipative) part of the conductivity,

\begin{equation}
D= \frac{1}{2}[\omega \sigma''(\omega)]_{\omega \to 0}
=\frac{1}{L} \left( \frac{1}{2} \langle  - T \rangle -
\sum_n p_n \sum_{m\neq n}
\frac{ \mid \langle n \mid  j \mid m \rangle \mid ^2}
{\epsilon_m - \epsilon_n} \right); 
\label{drude}
\end{equation}
\noindent
here $\langle T\rangle$ denotes the thermal expectation value of the kinetic
energy, generalizing the zero temperature expression to $T>0$ 
by considering a thermal average.
Thus, a finite Drude weight implies an ``ideal
conductor", a freely accelerating system.
The second definition of the Drude weight follows from the familiar 
optical sum-rule \cite{mald,bgr,ss} using eq.(\ref{econd}),
\begin{equation}
\int_{-\infty}^{+\infty} \sigma'(\omega) d\omega =\frac{\pi}{L}
\langle -T \rangle,
\label{osum}
\end{equation}

\noindent
with the average value of the kinetic energy replacing for nearest 
neighbor hopping tight binding
models the usual ratio of density over mass of the carriers for systems in
the continuum. 

At $T=0$ the Drude weight $D_0=D(T=0)$ is the
central quantity determining charge transport. As already
formulated by Kohn in a very physical way, 
$D_0$ can also be expressed directly as the sensitivity of the
ground state energy $\epsilon_0$ to an applied flux $\phi=eA$ ($e=1$),

\begin{equation}
D_0=\frac{1}{2L}\frac{\partial^2 \epsilon_0}{\partial
\phi^2} \bigl|_{\phi \rightarrow 0}. 
\label{stif0}
\end{equation}
For a clean system, since at $T=0$ there cannot be any dissipation, 
one expects that 
$\sigma_{reg}(\omega \to 0)=0$ and we have to deal with two fundamentally
different possibilities with respect to $D_0$: 

\begin{description}
\item{} $D_0>0$ is characteristic of a {\it conductor} 
or {\it metal},
\item{} $D_0=0$ characterizes an {\it insulator}. 
\end{description}
The insulating state can originate from a 
filled electron band (usual band insulator) or for a non-filled
band from electron correlations, that is the Mott-Hubbard mechanism; 
the latter situation is  of
interest here. Note, that the same criterion of the sensitivity to flux
has been applied to disordered systems, in the context of  
electron localization theory \cite{thouless}.

The theory of the metal-insulator transition solely due to Coulomb
repulsion (Mott transition) has been intensively investigated in the last
decades by analytical and numerical studies \cite{imada} of particular 
models of correlated electrons and it is one of the better understood parts 
of the physics of strongly correlated electrons. 

\bigskip
At finite temperatures, within the usual Boltzmann theory for weak 
electron scattering, the
relaxation time approximation represents well the low frequency
behavior,
\begin{equation}
\sigma(\omega)=\sigma_{dc}/(1+i\omega \tau),
\label{drudeform}
\end{equation}
where the relaxation time $\tau$ depends on the particular scattering
mechanism and is in general temperature dependent. In the following, we
consider only homogeneous systems without any disorder, so the
relevant processes in the solid state are electron-phonon scattering
and the electron-electron (Coulomb) repulsion. When the latter becomes
strong it is expected to dominate also the transport quantities. 

Even in a metal with $D_0>0$ it is not evident which is the relevant
scattering process determining $\tau(T)$ and $\sigma_{dc}(T)$. In the absence
of disorder and neglecting the electron-phonon coupling the standard
theory of purely electron-electron scattering would state that one
needs Umklapp scattering processes to obtain a finite $\tau$.  That is,
the relevant electron Hamiltonian includes the kinetic energy
$H_{kin}$, the lattice periodic potential $V$ and the
electron-electron interaction $H_{int}$,
 
\begin{equation} 
H= H_{kin} + V +H_{int}.
\end{equation}
Then, in general, 
the electronic current density $j$ is not conserved
in an Umklapp scattering process as the sum of ingoing electron momenta 
equals the sum of outgoing ones only up to a nonzero reciprocal vector $G$, 
$\sum_i k_i=mG$.
In other words, the noncommutativity of the current with the 
Hamiltonian, $[H,j]\neq 0$, leads to current relaxation and
thus, by the fluctuation-dissipation theorem, to dissipation. 
The interplay of $V$ and $H_{int}$, however, turns out to be 
fairly involved in the case of strong electron-electron repulsion. This
will become clear in examples of integrable tight binding models of
interacting systems that we will discuss below, 
which have anomalous (diverging) transport coefficients.

Experiments on many novel materials, - strange metals - with
correlated electrons, question the validity of the concept of a
current relaxation rate $1/\tau$. Prominent examples are the 
superconducting cuprates with very anisotropic, nearly planar,
transport \cite{imada} where the experimentally observed $\sigma(\omega)$ 
in the normal state can be phenomenologically described only by strongly 
frequency (and temperature $T$) dependent $\tau(\omega,T)$. 
The experiments on $\sigma(\omega)$ in quasi-1D systems are covered 
elsewhere \cite{degiorgi}.

\begin{figure}
\sidebyside{
\epsfig{file=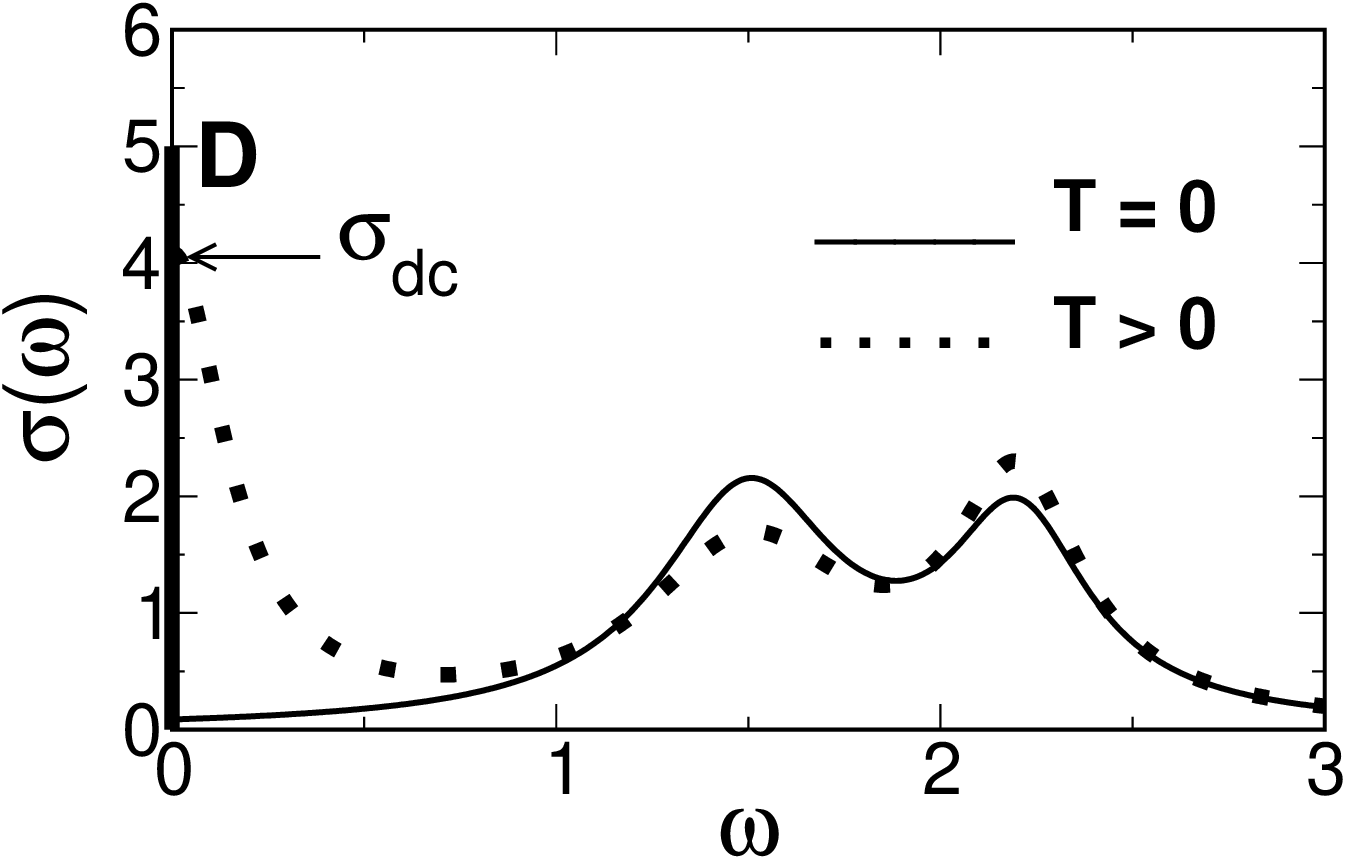, width=5.5cm,angle=0}
\caption{\label{f1}Schematic representation of the typical behavior of the 
conductivity for a clean metal.}}
{\epsfig{file=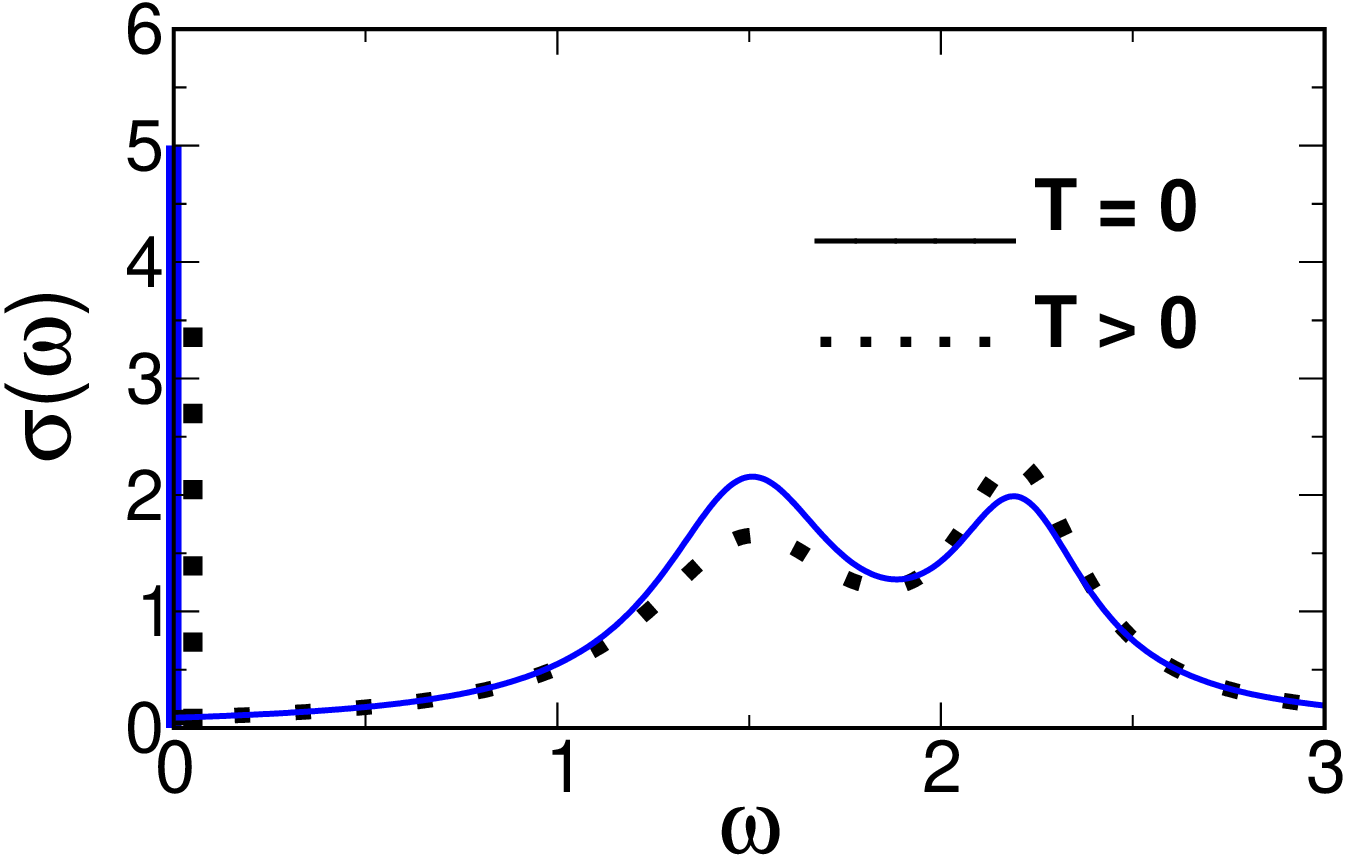, width=5.5cm,angle=0}
\caption{\label{f2}Illustration of the conductivity of a 
clean metal remaining an ideal conductor at finite temperatures.}}
\end{figure}

\bigskip
With this background, we will now discuss different possible scenaria  
for the behavior of the $T>0$ conductivity.
A clean metallic system at $T=0$  
is characterized by a $\delta-$function Drude peak and a finite frequency 
part that vanishes, typically with a power law dependence, implying 
zero $dc$ regular conductivity. 
In the common sense scenario, at finite temperatures the $\delta-$function 
broadens to a ``Drude peak"
of width inversely proportional to a characteristic scattering time and 
thus a finite $\omega\rightarrow 0$ limit implying a finite $dc$ conductivity.
The scattering mechanisms can be intrinsic, due to interactions, or extrinsic 
due to coupling to other excitations, phonons, magnons etc.
This typical behavior is shown in Fig. \ref{f1}.

Actually, for a finite size system (as often studied in numerical 
simulations) $D$ is nonzero even at finite $T$; it only goes to zero, 
typically exponentially fast, as the system size tends to infinity.
Physically, this expresses the situation where the thermal scattering 
length is less than the system size.

But it is also possible that constraints on the scattering mechanisms 
limit the current decay, so that the system 
remains an ideal conductor ($D>0$) even at $T>0$.
A schematic representation of a system remaining an ``ideal conductor" at 
finite $T$ is shown in Fig. \ref{f2}. 

\bigskip
In a system with disorder, $D$ vanishes even at zero 
temperature and the $dc$ residual conductivity is finite (provided the 
disorder is not strong enough to produce localization).

For an insulating system, e.g. due to interactions or the band 
structure as we discussed above, $D$ vanishes at zero 
temperature; in the conventional case, $D$ remains zero at $T>0$, while 
activated carriers scattered via different processes give rise 
to a finite $dc$ conductivity.
But it is also possible that $D$ becomes finite, turning a $T=0$ insulator to 
an ideal conductor; for instance, a system of independent particles 
(e.g. one described within a mean field theory scheme), 
insulating due to the band structure, turns to an ideal conductor at $T>0$. 
Finally, it is also conceivable that both $D$ and $\sigma_{dc}$ 
remain zero at $T>0$, a system that can be called an ``ideal insulator".

To the above scenaria we should add the possibility that the low 
frequency conductivity at finite temperatures is anomalous, e.g. diverging 
as a power law of the frequency, resulting in an infinite $dc$ 
conductivity. Actually, as we will discuss later (Discussion section), 
this kind of behavior is fairly common 
in classical one dimensional nonlinear systems.

\bigskip
Thus, the first step in characterizing a system is the evaluation of the 
Drude weight at $T=0$ in order to find out whether the system is 
conducting or insulating. The peculiarity that has recently 
been noticed is that most prototype models, 
assumed faithful representations of the physics 
of several quasi-one dimensional materials, 
have finite Drude weight also at finite temperatures 
(even $T\rightarrow\infty$), thus implying intrinsically ideal conductivity.  
In other words, interactions do not present a sufficient scattering 
mechanism to turn these systems into normal conductors.
This behavior is unlike the one observed in the higher dimensional 
version of the same models, that become normal conductors at finite 
temperatures \cite{tjpp}. 
This unconventional behavior has been attributed to the integrability 
of these models.

To evaluate the Drude weight is not an easy matter as, although frequency 
independent, it represents a transport property and thus it cannot be  
obtained via a thermodynamic derivative (e.g. of the free energy). 
Direct calculation using the optical sum 
rule eq.(\ref{osum})  is obviously involved requiring the value of 
all current matrix elements. A very convenient and physical formulation 
is the one by W. Kohn, eq.(\ref{stif0}), that generalized at finite 
temperatures \cite{czp} reads, 
\begin{equation}
D = \frac{1}{2L} \sum_n p_n
\frac {\partial ^2 \epsilon_n (\phi) }{\partial \phi^2}|_{\phi
\rightarrow 0} .
\label{dcurv}
\end{equation}

\noindent
By considering the change of the free energy as a function of flux 
(that vanishes in the thermodynamic limit as it is proportional to the 
susceptibility for 
persistent currents) we can also arrive at an expression for the 
Drude weight as the long time asymptotic value of current-current 
correlations \cite{znp},

\begin{equation}
D=\frac{\beta}{2L}\sum_n p_n \langle n| j |n\rangle ^2=
\frac{\beta}{2L} \langle j(t)j\rangle_{t\rightarrow \infty}\equiv
\beta C_{jj}.
\label{cjj}
\end{equation}

\noindent
As an example, for a 1D tight binding free spinless fermion system with nearest 
neighbor hopping $t$, the application of a flux $\phi$ 
modifies the single particle dispersion to $\epsilon_k=-2t\cos(k+\phi)$ 
giving,

\begin{eqnarray}
D_0=\frac{t}{\pi}\sin (\pi n)=
N(\epsilon_F) j_F^2\nonumber\\
D(T)\sim D_0-\frac{\pi t}{12}(\frac{T}{t})^2~~~(n=\frac{1}{2}).
\label{freed}
\end{eqnarray}

\noindent
Here, $n$ is the fermion density, $N(\epsilon_F)$ the density of 
states and $j_F$ the current at the Fermi energy.
Notice the quadratic decrease with temperature of the Drude 
weight that, as we will see later, it is generic even for interacting 
one dimensional fermionic systems out of half-filling.

\bigskip
In the recent literature, that we will discuss below,
the Drude weight of integrable systems is evaluated 
by the BA technique at zero or 
finite temperatures using the Kohn expression (\ref{dcurv}). 
The difficulty in this approach is the need for the 
estimation of finite size energy corrections of the order of $1/L$,  
a rather subtle procedure within this method.

Another approach, proved particularly efficient in establishing that systems 
with a finite Drude weight at finite temperature exist, uses an inequality 
proposed by Mazur \cite{mazur}. 
This inequality states that if a system is characterized by conservation laws
$Q_n$ then:

\begin{equation}
\lim_{{\cal T}\rightarrow \infty} \frac{1}{{\cal T}} \int_0^{\cal T} 
\langle A(t)A \rangle dt \geq \sum_n 
\frac{\langle A Q_n \rangle^2}{\langle Q_n^2 \rangle} .
\label{mazur}
\end{equation}
Here $\langle ~~~ \rangle$ denotes a thermodynamic average,
the sum is over a subset of conserved quantities ${Q_n}$ orthogonal
to each other in the sense 
$\langle Q_n Q_m \rangle=\langle Q_n^2\rangle\delta_{n,m}$, 
$A^{\dagger}=A$ and we take $\langle A\rangle=0$.

Thus, for time correlations $\langle A(t)A\rangle$ 
with non-singular low frequency
behavior we can obtain a bound for 
$C_{AA}=\lim_{t\rightarrow \infty} \langle A(t)A \rangle$,
\begin{equation}
C_{AA} \geq \sum_n \frac{\langle A Q_n\rangle ^2}{\langle Q_n^2\rangle } .
\label{mazur2}
\end{equation}

For integrable systems, such as the spin-1/2 Heisenberg or Hubbard model 
that are known to possess nontrivial conservation laws because of their 
integrability, useful bounds can be obtained by considering just the first 
non-trivial conservation law. We should stress however that this approach has 
not provided yet a complete picture of the Drude weight behavior as we 
will discuss below in concrete examples.

Finally, another argument relating the behavior of the Drude weight to the 
(non-) integrability of a model is by the use of 
Random Matrix Theory \cite{wilkinson,altshuler,czp}.
It is known that integrable systems are characterized by energy 
level crossings upon varying a parameter and so Poisson statistics 
in the energy level spacing; thus it can be argued that the typical 
value of diagonal current matrix elements (slope of energy levels with respect 
to an infinitesimal flux) is of the order of one, plausibly implying a 
finite Drude weight according to eq.(\ref{cjj}). 
On the contrary, nonintegrable systems, due to level repulsion, are 
described by Wigner or GOE statistics and thus the characteristic value 
of diagonal current matrix elements is of the 
order of $e^{-L}$ (inversely proportional to the density of many body states) 
implying now a vanishing Drude weight as $L\rightarrow \infty$. 

\bigskip
Besides electrical transport, the thermal conductivity of 1D systems
has recently attracted particular interest; within linear response
theory it is given by the analogous Green-Kubo formula expressed in terms 
of the energy current - energy current
dynamic correlation function,

\begin{equation}
\kappa(\omega)=\Re \frac{\beta}{i\omega} \chi_{j^Ej^E} (\omega).
\end{equation}

\noindent
Unlike the conductivity, there is no ``mechanical force" (as the flux $\phi$) 
that can be applied to the system in order to deduce expressions similar 
to the Drude weight, but the long time asymptotic value of energy 
current correlations has an analogous meaning. 

Finally, in magnetic systems, the ``spin conductivity"
(spin diffusion constant) can be probed, for 
instance, by NMR experiments that measure at high temperatures 
the Fourier transform of 
spin-spin autocorrelations at the Larmor frequency $\omega_N$,

\begin{equation} 
S(\omega_N)=\int_{-\infty}^{+\infty} dt 
\int dq e^{i\omega_N t} \langle S^z_{q}(t) S^z_{-q}\rangle.
\end{equation} 

\noindent
By using the continuity equation,

\begin{equation} 
\omega^2  \langle S^z_{q}S^z_{-q}\rangle_{\omega}
=q^2 \langle j^z_{q}j^z_{-q}\rangle_{\omega}
\end{equation} 

\noindent
for a system where the total spin $z-$component is conserved, 
the spin-spin dynamic correlations can be analyzed via the corresponding 
spin-current correlations in analogy to electrical transport \cite{nz}; 
the role of local charge is played by the $z-$ component of the local 
magnetization (see next section for a more detailed discussion on this 
point). 

\bigskip
We will now briefly discuss different methods, analytical and numerical, 
that are available for the study of finite temperature dynamic 
correlations in strongly interacting systems.
Among the analytical approaches that have been used for the study of 
transport and 
dynamic properties of 1D systems, each has its own advantages and drawbacks. 
The traditional memory function approach \cite{woelfle} provides a complete 
picture of the temperature/frequency dependence but it is a perturbative 
method based on the assumption of a regular relaxation behavior that might be 
dangerous in 1D systems. The high temperature moment expansion 
provides useful information on the possibility of anomalous transport 
but the extraction of transport coefficients is also based on the 
phenomenological assumption of regular, diffusive behavior \cite{moment}.
Progress in the exact evaluation of dynamic correlations 
in integrable systems has recently been achieved in the 
calculation of the Drude weight by the Bethe ansatz technique and of the 
frequency dependent conductivity by the form factor method. 
The Drude weight studies however are still controversial as they involve 
the calculation of finite size corrections, while the 
form factor approach has so far been limited to the calculation of 
zero temperature correlations and mostly in gapped systems. 
It is expected however that progress in BA techniques will 
provide a full picture of the dynamic properties of integrable systems. 
It is amusing to remark the paradoxical situation
where the only strongly correlated systems for which we can probably have 
a complete solution of their dynamics are the integrable ones, which however, 
exactly because of their integrability, show unconventional behavior.

\bigskip
Among numerical simulation techniques, the ED (exact diagonalization) 
provides exact answers over the full temperature/frequency range 
but of course only on finite size systems \cite{ftlm}. Due to the exponentially 
growing size of Hilbert space, this limits the size of systems that can be 
studied to only about 20 to 30 sites, depending on the complexity of the 
Hamiltonian. We should also remark that, in principle,  
the full excitation spectrum 
is required for the evaluation of finite temperature 
correlations 
\footnote{In a recent advance, finite temperature
dynamic correlations for a prototype
model have been successfully evaluated using only one quantum state
(microcanonical ensemble) \cite{micro}.}
Furthermore, the obtained frequency spectra are discrete, 
$\delta-$functions corresponding to transitions between energy
levels, so that some ad-hoc smoothing procedure is needed; this is 
particularly crucial in attempting to extract the low frequency behavior. 
Nevertheless, finite size scaling in 1D systems can provide very useful 
hints on the macroscopic behavior, particularly at high temperatures 
where all energy levels are involved. 
This regime is the most favorable in attempting to 
simulate the physical situation where the 
scattering length is less or comparable to the system size.

The Quantum Monte Carlo techniques allow the study of far larger systems 
and they provide directly the dynamic correlations at finite temperatures but 
in imaginary time \cite{jarrel}. 
By analytical continuation, using for instance the Maximum Entropy 
procedure, one is able in principle to extract the main features of the 
frequency dependence; experience shows however, that fine issues as  
the temperature dependence of the Drude weight or the presence of diffusive 
behavior which is a low frequency property, are difficult to establish 
reliably. 

Finally, the DMRG method that has 
been so successful in the study of ground state and thermodynamic properties 
of 1D systems, has only recently been extended to the reliable study of zero 
temperature conductivities in gapped systems \cite{white,eric}. 
At finite temperatures it is also possible to obtain very high accuracy data 
on autocorrelation functions 
in imaginary time by the use of the transfer matrix DMRG \cite{nwz}. 
However, similarly to QMC methods, it is very difficult to extract 
subtle information on the finite $T$ dynamics because of the extremely  
singular nature of analytic continuation that hides the useful information 
even for practically exact imaginary time data.

\section{Heisenberg model}

The prototype model for the description of localized magnetism is the 
Heisenberg model. For a one dimensional system the minimal Hamiltonian 
describing magnetic insulators is,
\begin{equation}
H=\sum_{l} h_l=J\sum_{l=1}^{L} (S_l^x S_{l+1}^x +
S_l^y S_{l+1}^y + \Delta S_l^z S_{l+1}^z)
\label{heis}
\end{equation}
where $S_l^{\alpha}~~(\alpha=x,y,z)$ are spin operators on site $l$ 
ranging from the most 
quantum case of spin S=1/2 to classical unit vectors. For S=1/2 the system 
is integrable by the Bethe ansatz method and its ground state, 
thermodynamic properties and elementary excitations have well been 
established \cite{korepin}. 
As a brief reminder to the discussion that follows, note that for $J>0$, 
$\Delta >0$ corresponds to an 
antiferromagnetic coupling while $\Delta <0$ to a ferromagnetic one;
a canonical transformation maps $H(\Delta)$ to $-H(-\Delta)$.
Further, the anisotropy parameter $\Delta$ describes two 
regimes, the ``easy-plane" for $|\Delta|<1$ or the ``easy-axis" for 
$|\Delta|>1$. The isotropic case, occuring in most materials for symmetry 
reasons, corresponds to $\Delta=1$. 
For $|\Delta|\le 1$ the system is gapless and 
characterized by a linear spectrum 
at low energies, while for $\Delta>1$ a gap opens;
in particular, at $\Delta=1$ the elementary excitation spectrum is 
described by the ``des Cloiseaux-Pearson" dispersion 
$\epsilon_q=\frac{J\pi}{2}|\sin q|$.
For $\Delta<-1$ there is a transition to a ferromagnetic ground state.

In general, other types of terms appear in the description of 
quasi-1D materials such as longer range or on site anisotropy interactions, 
but in this review we will focus on the prototype model eq.(\ref{heis}).

\bigskip
At this point we should mention that the spin-1/2 Heisenberg model is
equivalent to a model of interacting spinless fermions (the ``t-V" model)
obtained by a Jordan-Wigner transformation \cite{lsm};

\begin{equation}
H=(-t) \sum_{l=1}^L (c_{l}^{\dagger} c_{l+1} + h.c.)
+ V \sum_{l=1}^L (n_{l}-\frac{1}{2})(n_{l+1}-\frac{1}{2}) ,
\label{tv}
\end{equation}
where $c_{l}(c_{l}^{\dagger})$ denote annihilation (creation)
operators of spinless fermions at site $l$
and $n_{l}=c_{l}^{\dagger}c_{l}$.

The correspondence of parameters is $V/t=2\Delta$ and the opening of a gap
at $\Delta\ge 1$ corresponds to an interaction driven metal-insulator
(Mott-Hubbard type) transition.

\subsection{Currents and dynamic correlations}
Regarding the transport and dynamic properties of the Heisenberg model, 
three cases have mostly been discussed: the classical one, the spin S=1 
and the spin S=1/2; the S=1 case has been extensively analyzed by mapping its 
low energy physics to a field theory \cite{s1hald}, 
the nonlinear-$\sigma$ model (see section 5). 
In connection to experiment, the main issue is the diffusive vs. ballistic 
character of spin transport as probed for instance by NMR 
experiments and recently the contribution of magnetic excitations 
to the thermal conductivity of quasi-one dimensional materials \cite{ott}.

To discuss magnetic transport, we must first define the relevant spin 
$j^z$ and energy $j^E$ currents by the continuity equations of the 
corresponding local spin density $S^z_l$ (provided the total $S^z$ 
component is conserved) and local energy $h_l$;

\begin{equation}
S^z=\sum_l S_l^z,~~~~
\frac{\partial S_l^z}{\partial t}+\nabla j_l^z=0,
\label{contjs}
\end{equation}

\noindent
gives for the spin current, 
\begin{equation}
j^z=\sum_l j^z_l=J\sum_{l} (S_l^x S_{l+1}^y-S_l^y S_{l+1}^x).
\end{equation}

\noindent
Here and thereafter, $\nabla a_l=a_l - a_{l-1}$ denotes the discrete 
gradient of a local operator $a_l$.
In general ($\Delta\ne 0$) the spin current does not commute with the 
Hamiltonian, $[j^z,H]\neq 0$, so that nontrivial relaxation is expected and  
thus finite spin conductivity at $T > 0$. 

Similarly, the energy current $j^E$ is obtained by,
\begin{equation}
j^E=\sum_l j_l^E,~~~~
\frac{\partial h_l}{\partial t}+\nabla j_l^E=0,
\end{equation}

\begin{eqnarray}
j^E&=&J\sum_{l} 
          (S_{l-1}^x S_l^z S_{l+1}^y - S_{l-1}^y S_l^z S_{l+1}^x)
+\Delta   (S_{l-1}^y S_l^x S_{l+1}^z - S_{l-1}^z S_l^x S_{l+1}^y)
\nonumber\\
&+&\Delta (S_{l-1}^z S_l^y S_{l+1}^x - S_{l-1}^x S_l^y S_{l+1}^z)
\end{eqnarray}

\bigskip
We will now briefly comment on the framework for discussing spin dynamics and 
in particular how it is probed by NMR experiments.
According to the spin diffusion phenomenology (for a detailed 
description see ref. \cite{kadanoff}) 
when we consider the $(q,\omega)$ correlations of a conserved quantity 
$A=\sum_l A_l$, 
such as the magnetization or the energy, it is assumed that it will show a 
diffusive behavior in the long-time
$t\rightarrow\infty$, short wavelength $q\rightarrow 0$ regime 
\footnote{
This phenomenological statement goes under the name of Ohm's law in the 
context of electrical transport, Fourier's law for heat or Fick's law 
for diffusion.}. 
In the language of dynamic correlation function, diffusive behavior 
means that the time correlations decay as,
\begin{equation}
\langle \{A_l(t),A_{0}(0)\}\rangle =2\chi_A T\int \frac{dq}{2\pi}  
e^{iql-D_A q^2 |t|}
\label{diff}
\end{equation}
where $D_A$, $\chi_A$ are the corresponding diffusion constant and  
static susceptibility, respectively. 
For a 1D system, this behavior translates to a characteristic $1/\sqrt t$ 
dependence of the autocorrelation function.

Fourier transforming the above expression we obtain, 
\begin{equation}
S_{AA}(q,\omega)=\int_{-\infty}^{+\infty} dt~ e^{i\omega t} 
\frac{1}{2} \langle\{A_q(t),A_{-q}(0)\}\rangle
\sim\frac{\chi_A D_Aq^2}{(D_Aq^2)^2+\omega^2}.
\label{diffqw}
\end{equation}

\noindent
By using the continuity equation (\ref{contjs}), 
this Lorentzian form can be further 
modified to obtain the current-current correlation function,  
\begin{equation}
S_{j^Aj^A}(q,\omega)\sim\frac{\chi_A D_A\omega^2}{(D_Aq^2)^2+\omega^2}
\label{diffjj}
\end{equation}
which gives the diffusion constant $D_A$ by taking the 
$q\rightarrow 0$ limit first and then $\omega \rightarrow 0$.

On the other hand, a ballistic behavior is signaled by a $\delta-$function 
form, $S_{j^Aj^A}(q,\omega)\sim \delta(\omega-cq)$, where $c$ is a 
characteristic velocity of the excitations. This $\delta-$function peak 
moves to zero frequency as $q\rightarrow 0$ and its weight is proportional 
to the long time asymptotic of the current-current correlations 
\begin{equation}
C_{j^Aj^A}=S_{j^Aj^A}(q=0,t\rightarrow \infty).
\label{ball}
\end{equation}

\noindent
The above anticommutator correlations are related to the imaginary 
part of the susceptibility $\chi(q,\omega)$, 
that describes the dissipation, by,

\begin{equation}
S_{AA}(q,\omega)=\coth(\frac{\beta \omega}{2})\chi_{AA}''(q,\omega).
\label{saachi}
\end{equation}

In relation to the experimental study of spin dynamics, the NMR has developed 
to a very powerful tool; for instance, the $1/T_1$ relaxation time is directly 
related to the spin-spin autocorrelation by,

\begin{equation}
\frac{1}{T_1}\sim |A|^2 \int_{-\infty}^{+\infty} dt \cos(\omega_Nt) 
\langle \{S^z_l(t),S^z_l(0)\} \rangle
\label{t1}
\end{equation}

\noindent
where $|A|^2$ is the hyperfine coupling \cite{takigawa} and $\omega_N$ the 
Larmor frequency. Using the relation (\ref{saachi}), 
$1/T_1$ gives information (in the high 
temperature limit, $\beta \omega_N \rightarrow 0$) on $\chi''(q,\omega)$ 
as,

\begin{equation}
\frac{1}{T_1} \sim T|A|^2 \sum_q \frac{\chi''(q,\omega_N)}{\omega_N}.
\label{t1auto}
\end{equation}

\noindent
The diffusive behavior, characterized by the $1/\sqrt t$ decay of the spin 
correlations, is extracted in an NMR experiment by analyzing 
the $q\rightarrow 0$ contribution \cite{thurber}.  
It gives a $1/\sqrt \omega_N$ behavior that is detected as a 
$1/\sqrt H$ magnetic field dependence,

\begin{equation}
\frac{1}{T_1} \sim \frac{1}{\sqrt \omega_N} 
\sim \frac{T \chi(q=0)}{ \sqrt {D_s H}}, 
\label{t1chi}
\end{equation}

\noindent
considering that the Larmor frequency $\omega_N \sim H$, 
$D_s$ being the spin diffusion constant and $\chi(q=0)$ the static 
susceptibility.

\subsection{Spin and energy dynamics}
Returning now to the state of spin and energy dynamics, the classical 
Heisenberg model has been extensively studied by numerical simulations, 
the first studies dating from the 70's \cite{lurie}. 
Nevertheless, the issue of diffusive behavior (even at $T=\infty$ where 
most simulations are carried out) still seems not totally clear, 
the energy and spin showing distinctly different dynamics.
On the one hand, simulations clearly indicate that energy transport 
is diffusive \cite{reiter} but on the other hand, the decay of spin 
autocorrelations is probably inconsistent with the expected 
$1/\sqrt t$ law \cite{muller,reiter} exhibiting long-time tails.

On the other extreme, for the fully quantum spin S=1/2 model, the simplest 
case is the $\Delta=0$, so called XY limit. 
Here, the spin current commutes with the 
Hamiltonian resulting in ballistic transport; this can also be seen  
in the fermionic, t-V, version of model that corresponds to free spinless 
fermions  ($V/t=0$ in eq.(\ref{tv})) where now the charge current is conserved.
In the infinite temperature limit ($\beta=0$) the spin and energy 
autocorrelations can be calculated analytically using the Jordan-Wigner 
transformation and are of the form \cite{nie}:

\begin{equation}
\langle S^z_l(t) S^z_l \rangle = \frac{1}{4} \,J_0^2(Jt)
\label{xyszt}
\end{equation}
\begin{equation}
\langle h_l(t) h_l \rangle = \frac{J^2}{8} \,\left( J_0^2(Jt)+J_1^2(Jt) \right)
\label{xyet}
\end{equation}
which both behave as $1/t$ for $t\rightarrow\infty$, unlike the 
$1/\sqrt t$ form in the diffusion phenomenology 
($J_0, J_1$ are Bessel functions).
Actually the $\beta=0$ limit, often theoretically analyzed for simplicity, 
is not unrealistic as the magnetic exchange 
energy $J$ can be of the order of a few Kelvin in some materials. 

\bigskip
For $|\Delta|<1$ the Drude weight at $T=0$ has been calculated using 
the BA method \cite{haldxxz,ss} and is given by,

\begin{equation}
D_0=\frac{\pi}{8}\frac{\sin(\pi/\nu)}
{\frac{\pi}{\nu}(\pi-\frac{\pi}{\nu})},
\end{equation}

\noindent
where $\Delta=\cos(\pi/\nu)$ \footnote{The parametrization of $\Delta$ in terms 
of $\nu$ is common in the BA literature as the formulation greatly simplifies  
for $\nu=$integer.}. For $\Delta > 1$, $D(T=0)=0$ as the system 
is gapped.

At finite temperatures, several numerical and analytical studies 
indicate that for $|\Delta| < 1$ the spin transport is ballistic 
\cite{fmccoy,zp,nma,narozhny,carmelo}, in accord with the conjecture that this  
behavior is related to the integrability of the model \cite{mccoy,saito,czp}.
Pursuing this conjecture, one can attempt to use the Mazur 
inequality eq. (\ref{mazur2}) in order to obtain a bound on the Drude weight 
and thus establish that the transport is ballistic. Inspection of the known 
conservation laws for the Heisenberg model \cite{grab} shows that already 
the first nontrivial one, $Q_3$,  
has a physical meaning; it corresponds to the energy current, $Q_3=j^E$
and it can be used to establish a bound for $D$ \cite{znp}, 

\begin{equation}
D(T)\geq \frac{\beta}{2 L}\frac{\langle j^z Q_3\rangle^2}
{\langle Q_3^2\rangle}.
\label{dq3}
\end{equation}

\noindent 
This expression can be readily evaluated in the high temperature limit 
($\beta\rightarrow 0$),

\begin{equation}
D(T)\geq \frac{\beta}{2} \frac{8 \Delta^2 m^2 (1/4-m^2)}
{1+8\Delta^2(1/4+m^2)},~~~~~~~m=\langle S_l^z\rangle,
\end{equation}
where $m$ is equal to the magnetization density in the Heisenberg model or  
to $n-1/2$ in the equivalent fermionic $t-V$ model ($n$ is the 
density).
It establishes that ballistic transport is possible at 
all finite temperatures in the 
Heisenberg ($t-V$) model; notice however, that the right 
hand side vanishes for $m=0$, that corresponds to the specific case of 
the antiferromagnetic regime at zero magnetic field or to the $t-V$  model 
at half-filling. Of course this does not mean 
that $D$ is indeed zero in these cases as this relation provides only a bound. 
It should also be remarked that the obtained bound is proportional to 
$\Delta ^2$ and so we do not recover the simple result that $D(T)>0$ in the 
XY-limit. Furthermore, it can be shown, using a symmetry argument,  
that even by taking into account 
all conservation laws the bound remains zero at $m=0$ \cite{znp}.

A BA method based calculation of $D(T)$ for $|\Delta|<1$ was also 
performed \cite{z},  using a procedure proposed for the Hubbard 
model \cite{fk}, that relies upon a certain 
assumption on the flux dependence (see eq.(\ref{dcurv})) of bound state 
excitations (``rigid strings"). The resulting behavior is summarized in 
Figs. \ref{f3} and \ref{f4}.
From this analysis the following picture emerges: 

\begin{figure}[h]
\epsfig{file=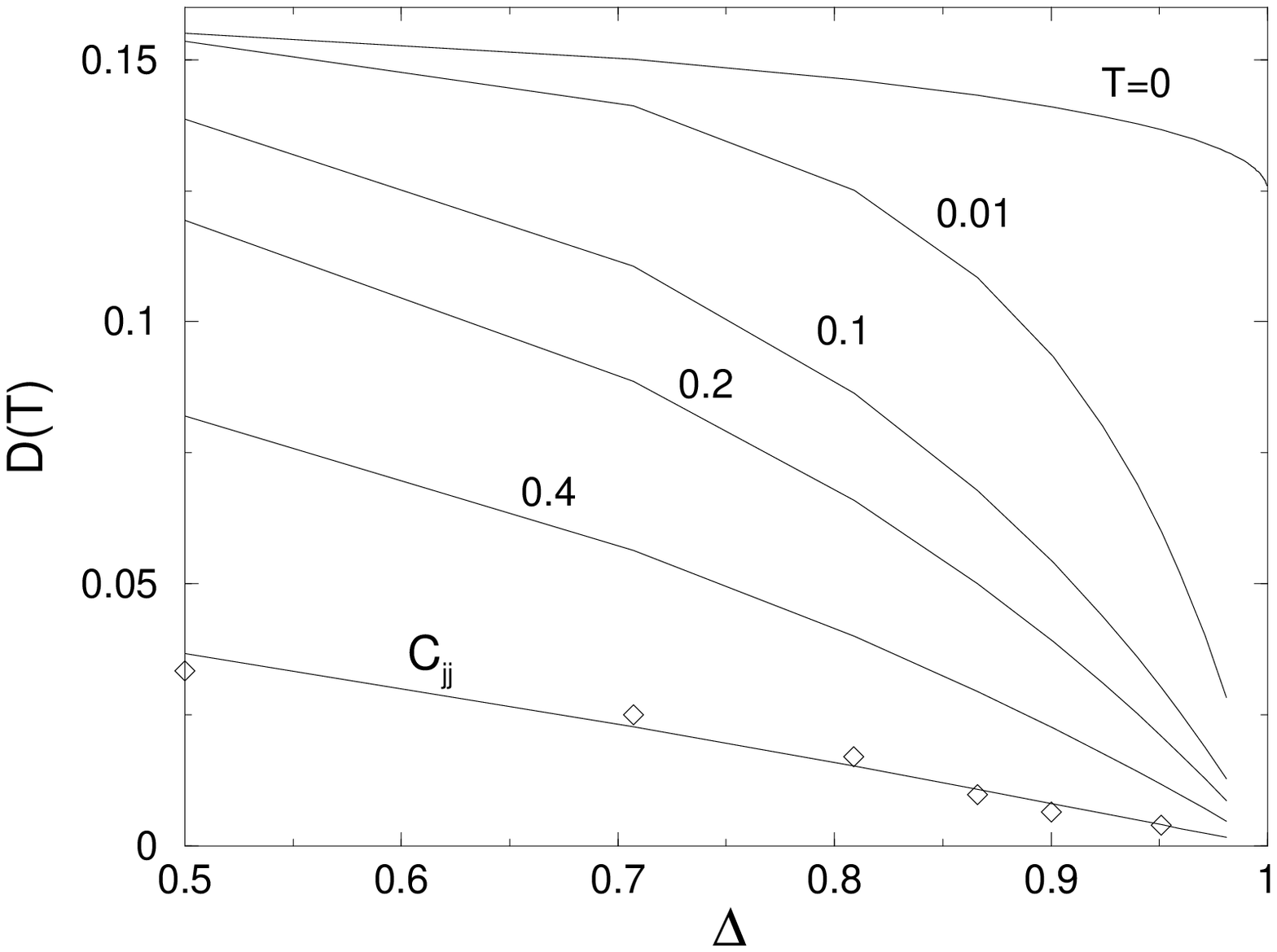, width=8cm, angle=0}
\vspace{-6cm}
\narrowcaption{$D(\Delta)$ at various temperatures. The lowest line is
the high temperature proportionality constant $C_{jj}=D/{\beta}$.
The symbols indicate exact diagonalization results \cite{nz}.}
\label{f3}
\end{figure}
\vspace{2cm}

\noindent
(i) at zero magnetization, in the easy plane antiferromagnetic regime 
($0<\Delta<1$), the Drude weight decreases at low 
temperatures as a power law 
$D(T)=D_0-{\rm const.}T^{\alpha},~~~\alpha=2/(\nu-1)$; 

\noindent
(ii) in the ferromagnetic regime, $-1<\Delta < 0$, $D(T)$ decreases 
quadratically with temperature (as in a noninteracting, XY-system);  

\noindent
(iii) the same low temperature quadratic behavior is true at any finite 
magnetization; 

\noindent
(iv) for $\beta\rightarrow 0$, $D(T)=\beta C_{jj}$
and it can be shown that $D(-\Delta)=D(\Delta)$
by applying a unitary transformation in the expression eq.(\ref{cjj});\\
a closed expression for $C_{jj}$ can be obtained by analytic calculations 
\cite{klumperpc}, 
$C_{jj}=(\pi/\nu-0.5 \sin(2 \pi/\nu))/(16\pi/\nu)$ for $|\Delta|<1$ 
while $C_{jj}=0$ for $\Delta>1$;

\noindent
(v) at the isotropic antiferromagnetic point ($\Delta=1$), $D(T)$
seems to vanish, implying non ballistic transport at all finite 
temperatures. 

\bigskip
\begin{figure}[h]
\epsfig{file=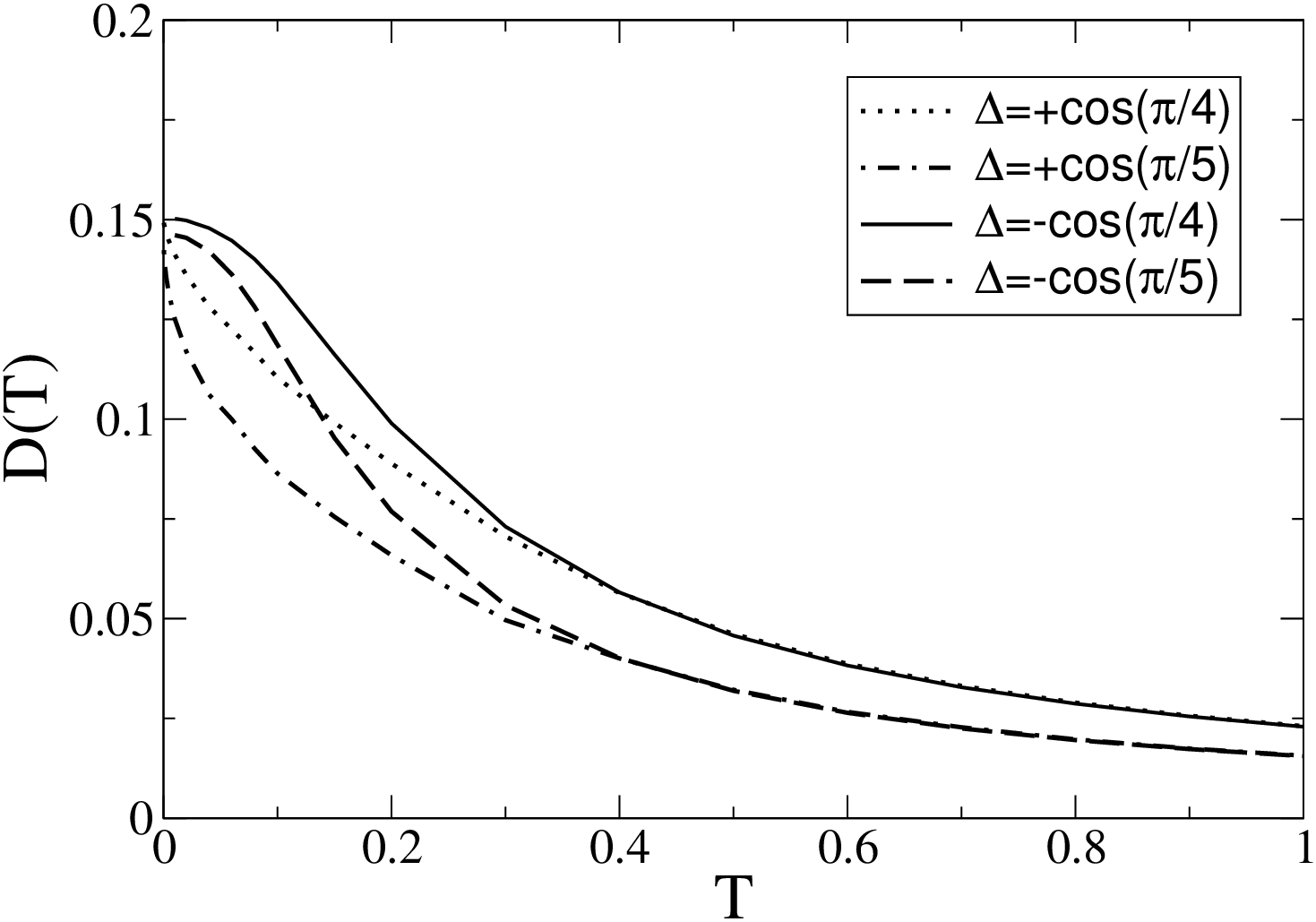,width=8cm, angle=0}
\vspace{-4cm}
\narrowcaption{Temperature dependence of the Drude 
weight $D$ vs. $T$ \cite{z}}
\label{f4}
\end{figure}
\vspace{2cm}

\noindent
This last result seems in accord with the most recent NMR data \cite{thurber}. 
Of course, the low frequency conductivity must also be 
examined in order to determine whether there is no anomalous behavior 
(e.g. power law divergence) that precludes a normal diffusive behavior;
such unconventional behavior is presently debated in classical
nonlinear 1D systems (see final section of Discussion).
It should not be surprising if future rigorous studies reveal 
that the isotropic Heisenberg exhibits a singular behavior, as it lies at the 
transition between a gapless and gapped phase.

In this context, we should also mention that the power law decrease of 
$D(T)$ for 
$0< \Delta < 1$ is not corroborated by recent QMC simulations \cite{gros}. 
The disagreement might be due either to the ``rigid string" assumption 
used in the BA analysis or to the very low temperatures,  
of the order of the energy level spacing, that are studied in the QMC 
simulations 
\footnote{Reliable results for the Drude weight can be obtained by QMC 
simulations only at low temperatures because a sufficiently 
fine spacing of Matsubara frequencies is required for the extrapolation 
to zero frequency.}.

Considering the limited results obtained so far 
using the Mazur inequality compared to the 
exact BA analysis, it remains an open question whether the behavior of the 
Drude weight can be fully accounted for solely by a proper consideration 
of conservation laws present in the Heisenberg model. 

\bigskip
For $\Delta >1$ numerical simulations \cite{zp} and analytical 
arguments \cite{carmelo} indicate that the Drude weight 
vanishes at all temperatures. In this regime, based on ED numerical 
simulations, it was proposed that a new phase might exist, an 
``ideal insulator", characterized by vanishing Drude weight and 
diffusion constant ($dc$ conductivity in the fermionic version). 
This conjecture remains presently still rather tentative, due to the small 
size of the systems that have been studied so far. 

On the other hand, a semiclassical field theory 
approach \cite{damle} concluded that gapped systems are diffusive. 
This approach is based on a mapping of the massive excitations to impenetrable 
classical particles of two or more charges (corresponding to different 
spin directions) that propagate diffusively (see section 5) and 
it has mostly been used for the analysis of gapped spin-1 systems. 

\bigskip
In parallel to these developments, the spin S=1/2 Heisenberg model was 
studied in the scaling limit using conformal invariance 
arguments \cite{schulz,sachdev}. This field theory approach amounts to 
considering a linearized spectrum and thus neglecting the effects of 
curvature, a point that we will further discuss below in section 5.
In particular, it was shown that 
the uniform dynamic susceptibility describes ballistic behavior,  
the corresponding $1/T_1$ relaxation time was evaluated and the theory 
was extensively compared to experimental data \cite{tak2}. Notice, however, 
that a later experimental 
NMR work \cite{thurber} concludes that the $q=0$ mode of spin transport 
is ballistic at the $T=0$ limit, but has a diffusion-like contribution 
at finite temperatures even for $T<<J$. 
We should remark that, over the years, the most common  
interpretation of NMR experiments was within the diffusion phenomenology, 
as for instance for the $S=5/2$ TMMC compound \cite{borsa}.

Finally, the finite ($q,\omega$) response functions of the S=1/2 model at 
$T=0$ were studied by the bosonization technique \cite{shankar} after mapping 
it to spinless fermions (eq.(\ref{tv})). For $\Delta < 1$, the conductivity 
shows the typical ballistic form; for $\Delta > 1$ it 
vanishes below the gap, showing a square-root frequency dependence above.

\bigskip
Turning now to energy transport, it is easy to see that 
the energy current is a conserved 
quantity \cite{nieth,znp} for all values of the anisotropy $\Delta$ 
implying that the currents do not decay and 
so the thermal conductivity is infinite. This peculiarity 
has also been 
noticed by an earlier analysis of moments at infinite temperature \cite{huber}.
So the quantity characterizing thermal transport is the equal time 
correlation $\langle j^Ej^E\rangle$ 
that represents the weight under the low frequency 
peak that will develop from the zero frequency $\delta-$function 
when a dissipative mechanism is introduced.
This picture is analogous to that of the electrical conductivity   
illustrated in Fig. 1.1. It implies 
that, given an estimate of the temperature dependence of the characteristic 
scattering time one is able to extract the value of the $dc$ 
thermal conductivity, further assuming some form (e.g. eq.(\ref{drudeform})) 
for the low frequency behavior.

\begin{figure}
\center{\epsfig{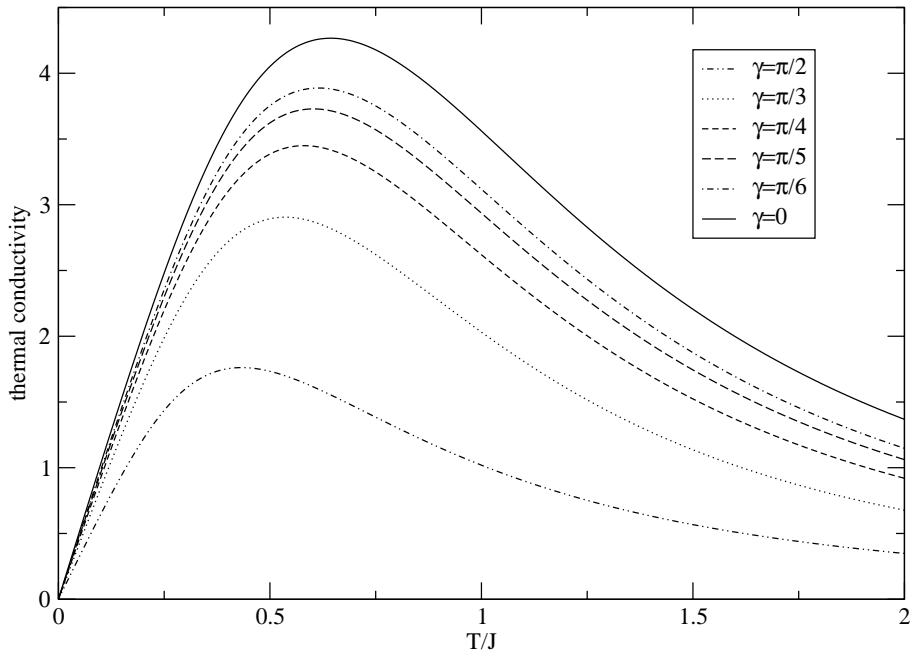}}
\caption{Thermal conductivity, $\langle j^E j^E \rangle$ 
in units of $J^2$, for various
anisotropy parameters $\Delta=\cos(\gamma)$ \cite{klumper}.}
\label{f5}
\end{figure}

This quantity has also recently been exactly calculated using the BA  
method \cite{klumper}; it is shown in Fig. \ref{f5}.

Furthermore, the experimental observation of 
unusually high thermal conductivity in ladder compounds \cite{ott} 
motivated the theoretical study of the thermal Drude weight in 1D anisotropic, 
frustrated and ladder spin-1/2 systems \cite{gros2,brenig}; 
the proposal of unconventional thermal transport in these systems 
is  still debated.

Finally, the S=1 Heisenberg chain shows a qualitatively different 
behavior characterized by the presence of an energy (Haldane) gap at 
low energies. The S=1 Heisenberg model is not integrable but 
the physics at low energies is usually mapped onto 
the quantum nonlinear-$\sigma$ 
model that is again an integrable system. The results known on this model will 
be briefly discussed in section 5 along with a semiclassical approach
to describe this type of gapped systems. The same low energy mapping is used 
for the analysis of ``ladder" compounds. 

\bigskip
As a guide to experimental investigations and theoretical studies, 
we can recapitulate the above discussion of the dynamics of the 
Heisenberg S=1/2 model as follows. 
It seems clear that ballistic behavior at all temperatures 
should be expected in the easy-plane 
regime and at all finite magnetizations, while the isotropic point is a 
subtle borderline case. The behavior in the easy-axis antiferromagnetic 
regime might be particularly interesting and it is not settled at the moment. 
Exceptionally high thermal conductivity should be expected in all regimes.

To complete the above picture, we should stress that not much is known on the 
low frequency behavior of the conductivities at finite temperature.
This leaves open the possibility of unconventional behavior, neither 
ballistic nor 
simple diffusive but one characterized by long time tails, giving rise to 
power law (or logarithmic) behavior at low frequencies. 

\section{Hubbard model}

The prototype model for the description of electron-electron correlations is 
the Hubbard model given by the Hamiltonian,

\begin{equation}
H=\sum_l h_l=
(-t) \sum_{\sigma,l} (c_{l\sigma}^{\dagger} c_{l+1 \sigma} + h.c.)
+ U \sum_{l} n_{l\uparrow} n_{l\downarrow}
\end{equation}

\noindent
where $c_{l\sigma}(c_{l\sigma}^{\dagger})$ are annihilation (creation)
operators of fermions with spin $\sigma=\uparrow, \downarrow$ at site $l$
and $n_{l\sigma}=c_{l\sigma}^{\dagger}c_{l\sigma}$.

At half-filling (n=1, 1 fermion per site) it describes a Mott-Hubbard 
insulator for any value of the repulsive interaction $U>0$, while it is a 
metal at any other filling.

The one dimensional Hubbard model is also integrable by the Bethe ansatz 
method and its phase diagram, elementary excitations, correlation 
functions have been extensively studied \cite{korepin,degkkk}. 

\subsection{Currents}
Similarly to the Heisenberg model, we can discuss the electrical, spin  
and thermal conductivity by defining the charge $j$, spin $j^s$ and 
energy $j^E$ currents from the respective continuity equations of the 
local particle density $n_l$, 

\begin{equation}
\frac{\partial n_l}{\partial t}+\nabla j_l=0,~~~
j=\sum_l j_l=\sum_{l\sigma} j_{l\sigma}=(-t)\sum_{\sigma,l}
(ic_{l\sigma}^{\dagger} c_{l+1 \sigma} + h.c.),
\end{equation}

\noindent
spin density $n_{l\uparrow}-n_{l\downarrow}$, 
\begin{equation}
\frac{\partial (n_{l\uparrow}-n_{l\downarrow})}{\partial t}+
\nabla j^s_l=0,~~~
j^s=\sum_l j^s_l=\sum_l j_{l\uparrow}-j_{l\downarrow}
\end{equation}

\noindent
and energy density $ h_l$,
\begin{eqnarray}
&&\frac{\partial h_l}{\partial t}+\nabla j^E_l=0,~~
j^E=\sum_{l,\sigma} j^E_{l\sigma}\\ 
j^E_{l\sigma}&=&(-t)^2(i c_{l+1\sigma}^{\dagger} c_{l-1\sigma} + h.c.)-
\frac{U}{2} j_{l,\sigma}
(n_{l,-\sigma}+n_{l+1,-\sigma}-1).\nonumber
\end{eqnarray}

\subsection{Electrical and thermal transport}
\begin{figure}
\center{\epsfig{file=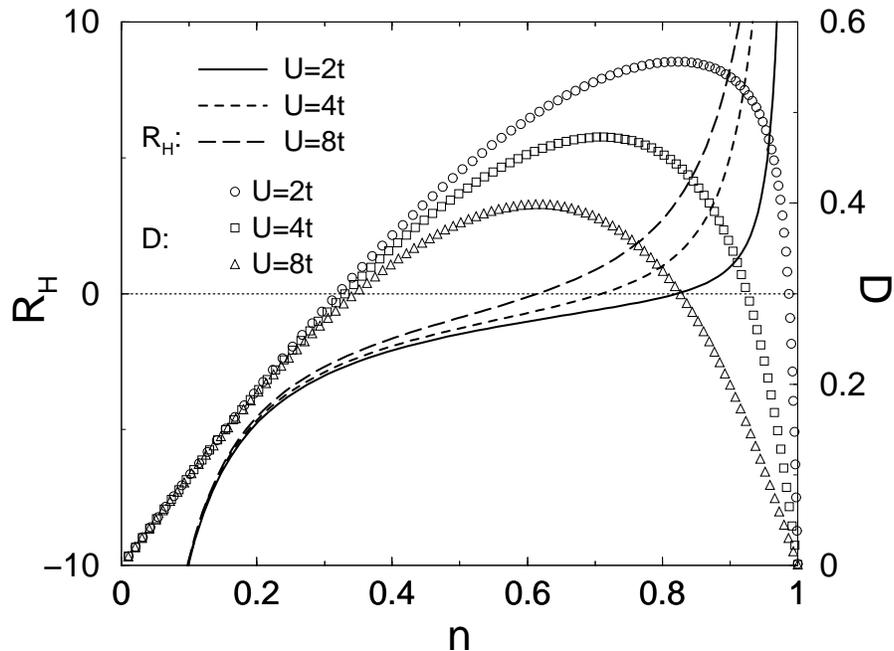,width=\textwidth,angle=0}}
\caption{Drude weight $D$ and $R_H$ for the quasi-1D Hubbard model from 
expression (\ref{naive}).}
\label{f6}
\end{figure}

With respect to the electrical conductivity   
the interaction $U$ and density dependence of the Drude weight $D$ 
at zero temperature has been established using the BA 
method \cite{haldd,schulzh,kawak} (see Fig. \ref{f6}).
There are two simple limits:

\noindent
(i) The free fermion case $U=0$ where $j$ is conserved and $D_0=
\frac{2t}{\pi}\sin \frac{\pi n}{2}$ where $n$ is the density of
fermions ($n=2 k_F/\pi$). Here $D_0$ vanishes for an empty band $n=0$
and a filled band $n=2$, being maximum at half filling, $n=1$. 

\noindent
(ii) Another simple limit is $U=\infty$. Since in this case the double
occupation of sites is forbidden, fermions behave effectively as
spinless fermions and the result is $D_0=\frac{t}{\pi}|\sin (\pi n)|$; 
here $D_0$ vanishes also at half filling.

Analytical results in 1D indicate that the $D_0=0$ value at half filling
persists in the Hubbard model for all $U>0$, whereby the density dependence 
$D(n)$ is between the limits $U=0$ and $U=\infty$. The insulating
state at half filling is a generic feature of a wider class of
1D models characterized by repulsive interactions, 
such as the $t$-$V$ model (discussed above), the $t$-$J$ model etc.

In Fig. \ref{f6}, along with the Drude weight, the zero 
temperature (ballistic) Hall constant $R_H$ of a quasi-1D system is also shown. 
According to a recent formulation \cite{znlp}, $R_H$ 
can be expressed in terms of the derivative of the Drude weight with 
respect to the density, 

\begin{equation}
R_H=-\frac{1}{D}\frac{\partial D}{\partial n}.
\label{naive}
\end{equation}

\noindent
The Hall constant is the classical way for determining the sign of the 
charge carriers. For a strictly one dimensional system of course it makes no 
sense to discuss the Hall effect; but if we consider a quasi-one 
dimensional system with interchain coupling characterized by a hopping 
$t'\rightarrow 0$, 
then within this formulation we recover a simple picture for the 
behavior of the sign of carriers as a function of interaction. 
In agreement with intuitive semiclassical arguments, the Hall constant 
behaves as $R_H\simeq -1/n$ at low densities changing to 
$R_H\simeq +1/\delta (\delta=1-n)$ near half-filling, with the turning point 
depending on the strength of the interaction $U$.
Notice that if $D \propto n$ with a small proportionality constant, 
that would be interpreted within 
a single particle picture as indicative of a large effective mass,  
then we would still find $R_H\simeq -1/n$. 
This observation might be relevant in the context of 
recent optical and Hall experiments \cite{mihaly,degiorgi} where a small Drude 
weight is observed although the Hall constant indicates a carrier density of 
order of one.

\begin{figure}
\center{\epsfig{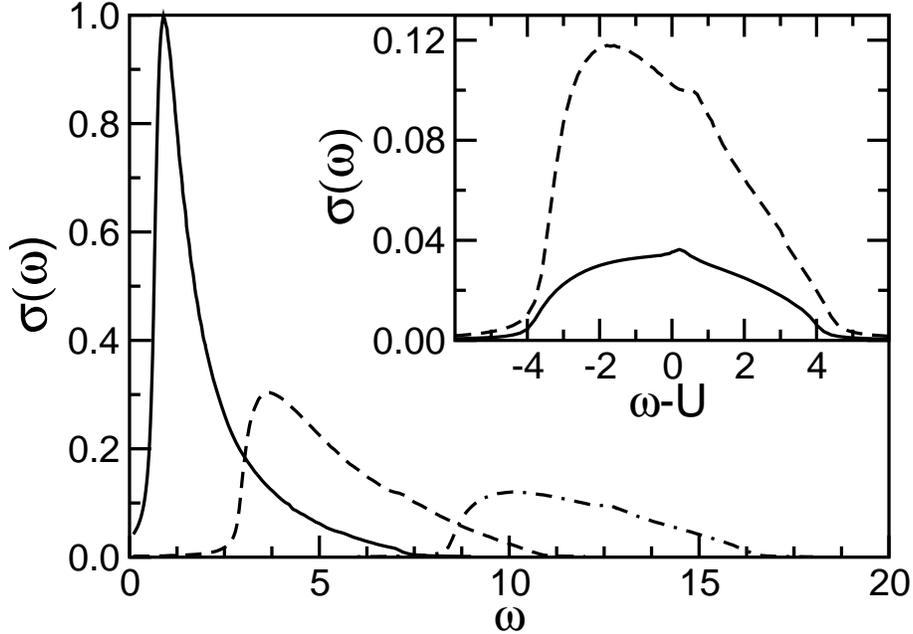}}
\caption{Optical conductivity at $T=0$ for $U/t=3,6,12$ (from left to right) 
calculated with DDMRG on a 128-site lattice \cite{jge}. Inset: 
$\sigma(\omega)$ for $U/t=12$ (dashed) and $40$ (solid) calculated on a 
64-site chain.}
\label{f7}
\end{figure}

Recently, using the form factor and DMRG methods the frequency 
dependence of the conductivity at half filling and at $T=0$ has also been  
studied \cite{jge} and is shown in Fig. \ref{f7}. 
The DMRG method provided the entire absorption spectrum for all but 
very small couplings where the field theoretical approach was used; the 
two methods are in excellent agreement in their common regime of 
applicability.
As expected, the Drude weight is zero, signaling an 
insulating state (for a detailed analysis of the scaling of $D$ with 
system size at and close to half-filling, see \cite{sm}) and the 
finite frequency conductivity vanishes up to the gap. 
Above the gap, a square root dependence is observed but not 
a divergence; this behavior is in contrast to that obtained by the 
Luttinger liquid method \cite{giam} and it is typical 
of a Peierls (band) insulator where a divergence occurs. 
This absence of a singularity is also in agreement with a rigorous analysis 
of the sine-Gordon (sG) field theory (see section 5), the 
generic low energy effective model for the description of a Mott-Hubbard 
insulator.
 
To complete the zero temperature picture, the frequency dependent 
conductivity of the Hubbard model out of half-filling has been studied 
using results from the BA method and symmetries \cite{carmelohub}. 
A broad absorption band was found separated from the Drude peak at $\omega=0$ 
by a pseudogap; this pseudo-gap behavior is in contrast to the $\omega^3$ 
dependence found within the Luttinger liquid analysis \cite{giam}.

\bigskip
Again, at all finite temperatures the transport is ballistic 
characterized by a finite Drude weight. In an identical way to the 
Heisenberg model, this can easily be established 
by the Mazur inequality 
using the first nontrivial conservation law $Q_3$. For the Hubbard model 
$Q_3$ differs from the energy current $j^E$ by the replacement of 
$U$ by $U/2$ \cite{znp}.
Evaluating $\langle j Q_3\rangle^2/\langle Q_3^2\rangle$ 
for $\beta \rightarrow 0$ 
we obtain,

\begin{equation}
D(T) \geq \frac{\beta}{2L} \frac{\langle j Q_3\rangle^2}
{\langle Q_3^2\rangle }=\frac{\beta}{2}
\frac
{[U\sum_{\sigma}2n_{\sigma}(1-n_{\sigma})(2n_{-\sigma}-1)]^2}
{\sum_{\sigma} 2n_{\sigma}(1-n_{\sigma})[1+U^2(2n_{-\sigma}^2-
2n_{-\sigma}+1)]}\nonumber,
\label{dhubb}
\end{equation}
where $n_{\sigma}$ are the densities of $\sigma=\uparrow,\downarrow$
fermions.

By inspection we can again see that from this inequality 
we cannot obtain a finite bound for $D(T)$ for 
$n_{\uparrow}+n_{\downarrow}=1$. 
Nevertheless, a full BA calculation \cite{fk} seems to show that the 
Drude weight at half-filling is exponentially activated 
$D(T)\sim \frac{1}{\sqrt T}  e^{-E_{gap}/T}$ at low temperatures and decreases 
as $T^2$ out of half-filling. Thus the zero temperature insulator turns to 
an ideal conductor at finite temperatures. 
Notice that this behavior is different from 
the one in the Heisenberg (or ``t-V") model in the gapped phase ($\Delta > 1$) 
where the Drude weight seems to vanish at all finite temperatures. 
We can conjecture that this distinct behavior of insulating phases 
can be understood in the framework of the corresponding low energy  
sine-Gordon field theory as these two models map to different parameter 
regimes of the sG model \cite{shankar,emery}.
A very similar calculation, using the Mazur inequality, can also 
be carried out for the long time asymptotics of the spin current, $j^s$, 
correlations. It gives a finite bound, and thus ballistic spin transport  
for $n_{\uparrow}-n_{\downarrow}\ne 0$; no BA calculation has so 
far been performed for the spin conductivity.

\bigskip
On the thermal conductivity we find similar results, namely a finite 
value on the long time decay of energy current correlations, which 
can readily be evaluated for $\beta\rightarrow 0$ \cite{znp},

\begin{eqnarray}
\lim_{t\rightarrow\infty} \langle j^E(t)j^E\rangle=C_{j^E j^E}  \geq
\frac{\langle j^E Q_3\rangle^2}{\langle Q_3^2\rangle}.
\label{kappah}
\end{eqnarray}

Again this inequality gives a finite bound for a system out of half-filling 
as long as $n_{\uparrow}+n_{\downarrow} \ne 0$ and this for 
any magnetization.
For this model the actual temperature dependence of 
$C_{j^Ej^E}=\lim_{t\rightarrow\infty} \langle j^E(t)j^E\rangle
=C_{j^E j^E}$ has not yet been evaluated.
Finally, the low temperature thermoelectric power was studied 
using the Bethe ansatz picture for the charge (holons) and spin (spinons) 
excitations \cite{staff}. The resulting sign of the thermopower close 
to the Mott-Hubbard insulating phase is 
consistent with the one derived from the Hall constant above, 
$S\sim sign(1-n)T|m^*|/|1-n|^2$.

In summary, we have shown that the prototype model for describing electron 
correlations in one dimensional systems, the 
Hubbard model, shows  unconventional, ballistic charge, spin and thermal 
transport at all finite temperatures. 
Of course real quasi-one dimensional materials 
are presumably characterized by longer range than the Hubbard $U$ 
interactions. So, although the above picture should be taken into account 
in the interpretation of experiments, (quasi-) one dimensional magnetic 
compounds might presently appear as better candidates for the experimental 
observation of these effects.
Theoretically, the full frequency dependence of the conductivities 
at finite temperatures remains to be established.

\section{Effective field theories}

An alternative to analyzing the transport of quasi-one dimensional 
materials within microscopic models, as
described in previous sections, is to approach the problem within 
effective low energy models for interacting
electrons, i.e. starting with the Luttinger liquid Hamiltonian. This
path is very attractive since it represents the counterpart of the
usual Landau phenomenological approach to Fermi liquid in higher-D
electronic systems. It should be pointed out that even in a 3D system
the continuum field theory is not enough to describe a current decay
and Umklapp processes are finally responsible for a finite intrinsic
resistivity $\rho(T) \propto T^2$ \cite{abri}.

In an effective (low energy) field theoretical model for 1D
interacting electrons the band dispersion around the Fermi momenta $k=\pm
k_F$ is linearized and left- and right- moving excitations are
defined. Apart from Umklapp terms, the model of interacting fermions
can then be mapped onto the well known Luttinger liquid 
Hamiltonian \cite{schoen,emery}
and analyzed via the bosonization representation. In particular one
obtains for the charge sector, 
\begin{equation}
H_{0}= \frac{1}{2\pi} \int dx \Bigl[ u_\rho K_\rho 
(\pi \Pi_\rho)^2 + \frac{u_\rho}{ K_\rho} (\partial_x \phi_\rho)^2
\Bigr],
\end{equation}
where the charge density is $\rho(x)=\partial_x \phi_\rho$ and
$\Pi_\rho$ is the conjugate momentum to $\phi_\rho$. Interactions
appear only via the velocity parameter $u_\rho$ and Luttinger exponent
$K_\rho$. The charge current in such a Luttinger model, $j=\sqrt 2
u_\rho K_\rho \Pi_\rho$, is clearly conserved in the absence of
additional terms.

Umklapp terms can as well be represented with boson operators,  
\begin{equation}
H_{\frac{1}{2m}}=
g_{\frac{1}{2m}} \int dx \cos(m\sqrt{8} \phi_\rho(x) +\delta x),
\label{sg}
\end{equation}
where $m$ is the commensurability parameter ($m=1$ at half-filling - 
one particle per site, $m=2$ for quarter filling - one particle for 
two sites etc)
and $\delta$ the doping deviation from the commensurate filling. 
In principle, the mapping of a particular (tight binding) microscopic model
onto a field theory model, e.g. via perturbation theory,
generates terms $H_m$ with arbitrary $m$.  While Umklapp terms 
are irrelevant in the sense of universal scaling of the static properties, 
they appear to be crucial for transport. They drive a metal at
half-filling to an insulator, while at an arbitrary (incommensurate)
filling they should cause a finite resistivity since the current is not
conserved any more (for an overview of the transport properties emerging 
within the Luttinger liquid picture see \cite{giamr}).

However, the proper treatment of transport within the Luttinger
picture in the presence of Umklapp processes is quite involved and even
controversial. Giamarchi \cite{giam} first calculated the effect of
Umklapp scattering within lowest order perturbation theory for the memory
function $M(\omega)$; he thus determined the low-$\omega$ behavior of the
dynamical conductivity $\sigma(\omega) \propto 1/(\omega+M(\omega))$ 
that yielded a nonzero finite temperature conductivity.
At the same time he realized, by using the Luther-Emery method, that the 
Umklapp term can be absorbed in the Hamiltonian in such a way as to conserve 
the current 
and pointed out the possibility of infinite $dc$ conductivity even in the 
presence of Umklapp. A similar lowest-order analysis \cite{gm} for
general commensurate filling predicts at $T=0$ that 
$\sigma(\omega) \propto \omega^{\nu-2}$ and the resistivity 
$\rho(T) \propto T^\nu$ with $\nu = 4n^2 K_{\rho} -3$.
On the other hand, Rosch and Andrei \cite{ra} pointed out that even in
the presence of general Umklapp terms there exist particular
operators, linear combinations of the translation operator and
number difference between left- and right- moving electrons, which are
conserved. Since in general such operators have a nonvanishing overlap with
the current operator $j$, this leads to finite $D(T>0)>0$ if only one
Umklapp term is considered. At least the interplay of two noncommuting
Umklapp processes is needed to yield a finite resistivity $\rho(T>0)>0$.

From a different perspective Ogata and Anderson \cite{pwa} argued that 
because of 
spin-charge separation in 1D systems an effect analogous to phonon drag 
(in this case spinon-holon drag) appears that leads to a finite dissipation. 
Using a Landauer like semi-phenomenological approach they concluded the 
existence of a linear-T resistivity and linear frequency dependence of 
the optical conductivity.

\bigskip
The bosonization of the Luttinger liquid model leads \cite{emery} 
to the quantum sine-Gordon model (eq.(\ref{sg})) which is an integrable 
system and has extensively been 
studied as a prototype nonlinear quantum (or classical) field theory.
It is the generic field theory for describing the low energy properties 
of one dimensional Mott insulators.
The thermodynamic properties and excitation spectrum consisting of 
solitons/antisolitons and breather states have been established 
by semiclassical and BA techniques \cite{korepin}. 
Presently, there is an effort to determine the transport properties of this 
model rigorously. In particular, the frequency dependence of the 
zero temperature conductivity in the commensurate (insulating) phase, 
zero soliton sector, 
has been evaluated using the form factor approach \cite{contro}. The main 
result is that the square root singularity at the optical gap, 
characteristic of band insulators, is generally absent and appears only at the 
Luther-Emery point; furthermore, the perturbative result \cite{giam} 
is recovered only at relatively high frequencies.
Besides these studies, 
the Drude weight and optical response near the metal-insulator transition, 
in the incommensurate phase at zero temperature, have also been studied by  
Bethe ansatz \cite{papa} and semiclassical methods \cite{luther}. 
Still, a rigorous evaluation of the 
Drude weight and frequency dependence of the conductivity at finite 
temperatures is missing; nevertheless, we can plausibly argue that because 
of the integrability of the sine-Gordon model, it will turn out 
that also this model describes an ideal conductor at least over 
some interaction range. Thus, it might remain an open question which scattering 
processes and/or band curvature must be taken into account 
in order to recover a normal, diffusive behavior at finite temperatures. 

\bigskip
Finally, it is well known \cite{rice} that the spectrum of integer spin 
and even-leg ladder systems  is gapped and that the low 
energy physics is described by the one-dimensional quantum $O(3)$ 
nonlinear sigma model \cite{s1hald}. 

In imaginary time $\tau$ the action at inverse temperature $\beta$ is given 
by 
\begin{equation}
S=\frac{c}{2g}\int_0^{\beta} d\tau \Bigl[ (\partial_x n_{\alpha})^2 +
\frac{1}{c^2} (\partial_{\tau} n_{\alpha})^2 \Bigr],
\end{equation}
where $x$ is the spatial coordinate, $c$ a characteristic velocity,  
$\alpha=1,2,3$ is an $O(3)$ vector index 
and $n_{\alpha}(x,\tau)$ a unit vector field $n^2_{\alpha}(x,\tau)=1$.

In a series of works, Sachdev and collaborators 
\cite{nlsm,sbook,buragohain} developed a picture of the 
low and intermediate temperature spin dynamics based on the idea that 
the spin excitations can be mapped to an integrable model describing 
a classical gas of impenetrable 
particles (of a certain number of species depending on the spin), a problem 
that can be treated analytically. Within this framework they have 
extensively analyzed NMR experiments on S=1 compounds \cite{takigawa1} 
and they concluded that these systems behave diffusively.
In contrast to this semiclassical approach, using the Bethe ansatz 
solution of the quantum nonlinear$-\sigma$ model \cite{wieg}, 
Fujimoto \cite{fuj} found a finite Drude weight, exponentially activated 
with temperature, and he thus 
concluded that the spin transport at finite temperatures is ballistic.
The origin of this discrepancy is not clear at the moment and can be due 
either to a subtle role of quantum effects on the dynamics that is 
neglected in the semiclassical approach or to 
a particular limiting procedure (the magnetic field going to zero) 
in the BA solution.

\section{Discussion}

We hope that the above presentation demonstrated that the transport theory of 
one dimensional quantum systems is a rapidly progressing field, 
fueled by both theoretical and experimental developments.
Still, on the question, what is the finite temperature conductivity of 
bulk electronic or magnetic systems described by strongly 
interacting one dimensional Hamiltonians,  
it is fair to say that no definite answer has so far emerged nor there is a  
clear picture of the relevant scattering mechanisms. 

In this context, it is interesting and instructive to draw an analogy with 
the development of the respective field in classical physics,  
namely the finite temperature transport in one dimensional nonlinear 
systems. Interestingly, in this domain we are also witnessing a 
flurry of activity after several decades of studies. Again, the issue of 
ballistic versus diffusive (usually energy) transport in 
a variety of models and the necessary ingredients for observing 
normal behavior is sharply debated \cite{leb,livi}. Similarly to the quantum 
systems, numerical simulations are intensely employed along with 
analytical approaches and discussions on the conceptual foundations 
of transport theory.

\bigskip
For quantum systems it is 
reasonable to expect that the finite temperature transport properties of
integrable models will, in the near future, be amenable to rigorous
analysis by mathematical techniques, for instance in the framework 
of the Bethe ansatz method. At the same time, as we mentioned earlier, 
it is amusing to notice that 
the integrable systems that we can exactly analyze, present 
singular transport properties presumably exactly because of their 
integrability. 

To obtain normal behavior, it is reasonable to invoke perturbations 
destroying the integrability of the model, as for instance longer range 
interactions, interchain coupling, coupling to phonons, disorder etc.
In this scenario, it is then necessary to find ways to study the effect of 
perturbations around an integrable system and in particular to determine 
the vicinity in parameter space around the singular-integrable point 
where unconventional transport can be detected.
This issue is also extensively studied in classical systems as it is 
the most relevant in the interpretation of experiments 
and in estimating the prospects for technological realizations.
It is worth keeping in mind the possibility that integrable interactions 
actually render the system more immune to perturbations, an effect 
well known and exploited in classical nonlinear systems \cite{osol}.

Related to this line of argument is the following question.  
If integrable models show ballistic transport and low energy effective 
theories like the sine-Gordon model are also integrable, then which mechanisms 
are necessary to obtain dissipative behavior ?

Of course it is also possible that 
the conventional picture according to which only integrable systems show 
ballistic transport might well be challenged.  
One dimensional nonintegrable quantum 
systems could also show singular transport in the form either of a finite 
Drude weight or low frequency anomalies. This behavior has been observed in 
classical nonintegrable nonlinear systems where the current 
correlations decay to zero in the long time limit but too slowly, so that 
the integral over time (giving the $dc$ conductivity) diverges.
The opposite behavior might also be realized, namely that integrable 
quantum systems 
show normal diffusive transport in some region of 
interaction parameter space (this possibility was raised in the case of  
gapped systems as the easy-axis spin 1/2 Heisenberg model or the 
quantum nonlinear$-\sigma$ model, see section 5).
Furthermore, the issue of the crossover of the dynamics between quantum 
and classical systems has, at the moment, very little been explored and in 
particular the question whether quantum fluctuations might stabilize 
ballistic transport behavior.

\bigskip
To address all the above open issues there is a clear need for 
the development of reliable analytical and numerical simulation 
techniques (as the DMRG or QMC) to tackle the evaluation of dynamic 
correlations at low temperatures. 
In particular, progress is needed to include the coupling between  
the different, magnetic, electronic and phononic, excitations.

\bigskip
In summary, one of the most fascinating aspects in this field is to 
understand the extent to which the so successful physics, experimental 
and technological realizations of classical (integrable) 
nonlinear systems can be carried over to the quantum world of many 
body (quasi-) one dimensional electronic or magnetic strongly interacting 
systems. This effort is accompanied by the experimental  
challenge to synthesize novel materials/systems that realize this physics.

\begin{acknowledgments}
It is a pleasure to acknowledge discussions over the last few years on this 
problem  with many colleagues, in particular 
H. Castella, 
F. Naef, 
A. Kl\"umper, 
C. Gros, 
A. Rosch,
D.Baeriswyl, 
H.R. Ott,
H. Beck, 
T. Giamarchi,
M. Long,  
N. Papanicolaou and a careful reading of the manuscript by D. Baeriswyl.
This work was supported by the Swiss National Foundation, the University of  
Fribourg, the University of Neuch\^atel and the EPFL through its 
Academic Guests Program.
\end{acknowledgments}

\end{document}